# Origin of High Temperature Oxidation Resistance of Ti-Al-Ta-N Coatings

R. Hollerweger[1], H. Riedl[1], J. Paulitsch[2], M. Arndt[3], R. Rachbauer[3], P. Polcik[4], S. Primig[5], P. H. Mayrhofer[1,2]

[1] Christian Doppler Laboratory for Application Oriented Coating Development at the Institute of Materials Science and Technology, Vienna University of Technology, 1040 Vienna, Austria

[2] Institute of Materials Science and Technology, Vienna University of Technology, 1040 Vienna, Austria

[3] Oerlikon Balzers Coating AG, 9469 Balzers, Principality of Liechtenstein

[4] Plansee Composite Materials GmbH, 86983 Lechbruck am See, Germany

[5] Department of Physical Metallurgy and Materials Testing Montanuniversität Leoben, A-8700 Leoben, Austria

## Abstract

Alloying Ti-Al-N coatings with Ta has proven to enhance their hardness, thermal stability, and oxidation resistance. However, especially for arc-evaporated Ti-Al-Ta-N coatings only limited information on the detailed influence of the elements on various properties is available. Therefore, we have developed arc-evaporated $Ti_{1-x-y}Al_xTa_yN$ coatings with various Al (x = 0.50 – 0.65) and Ta (y = 0.00 – 0.15) contents. While the thermal stability of our coatings during annealing in inert He atmosphere increases with increasing Ta content, best results are obtained for specific Ta-Al ratios during oxidation. Single phase cubic $Ti_{0.32}Al_{0.60}Ta_{0.08}N$ yields a mass-gain of only ~5 % after 5 h at 950 °C in synthetic air, whereas $Ti_{0.35}Al_{0.65}N$ is completely oxidized after 15 min. This is in part based on the suppressed anatase and direct rutile $TiO_2$ formation at a defined Ta-Al content. Consequently, the anatase-to-rutile transformation, generally observed for $Ti_{1-x}Al_xN$, is absent. This reduces the generation of pores and cracks within the oxide scale and especially at the nitride-oxide interface, leading to the formation of a protective rutile and corundum based oxide scale. This is also reflected in the pronounced decrease in activation energy for the protective scale formation from 232 kJ/mol for $Ti_{0.35}Al_{0.65}N$ down to 14.5 kJ/mol for $Ti_{0.32}Al_{0.60}Ta_{0.08}N$. Based on our results we can conclude that especially phase transformations within the oxide scale need to be suppressed, as the connected volume changes lead to the formation of cracks and pores.

## 1. Introduction

The development of hard protective coatings with increased thermal stability and oxidation resistance is in the main focus of several investigations [1–4]. Transition metal nitrides like TiN were one of the first industrially applied hard coatings. Even though TiN exhibits an excellent thermal stability, its mechanical properties at high temperatures as well as its oxidation resistance at temperatures higher than 500 °C is limited. Therefore, Münz [5] proposed to alloy TiN with aluminum to form a ternary $Ti_{1-x}Al_xN$ and thus to enhance its mechanical properties and oxidation resistance and hence to increase the lifetime of a protected tool.

As industrial demands on cutting inserts, mills and drills are continuously increasing, a further optimization of such protective coatings is necessary. Critical prerequisites are the phase stability of the nitride coating itself and a good oxidation resistance. Depending on the Al content, the supersaturated $Ti_{1-x-y}Al_xN$ phase starts to decompose at around 800 °C into isostructural face centered cubic (fcc) TiN and fcc-AlN. The latter is a metastable (high pressure) phase of AlN, which further transforms into hexagonal close packed (hcp, wurtzite type) AlN with a pronounced volume increase of ~24 % [6,7]. The thereby obtained dual phase structure (fcc-TiN and hcp-AlN) of the coating and the possibly generated micro-cracks not just result in a massive mechanical attack of the coating itself, but also leads to the formation of additional diffusion paths, decreasing the oxidation resistance [1,8,9]. Therefore, Holec et al. [10,11], Rachbauer et al. [12–15], and several other authors [16–22] have intensely studied the effect of trivalent, tetravalent and pentavalent alloying elements (X) on the thermal stability of sputter deposited $Ti_{1-x-y}Al_xX_yN$ coatings. These investigations showed that especially Ta (a pentavalent element) results in a pronounced increase in thermal stability of $Ti_{1-x-y}Al_xX_yN$ by shifting the onset of the nitride phase decomposition by ~200 °C to higher temperatures. The main reason for the increased thermal stability is based on the reduced chemical driving force for decomposition and retarded decomposition processes with increasing Ta content [11]. Additionally, alloying Ta to $Ti_{1-x}Al_xN$ is also very beneficial for increasing the oxidation resistance [15]. The increased oxidation resistance can only partly be understood by the reduced oxygen defects within rutile $TiO_2$ through alloying with pentavalent Ta [23], and the suggested promotion of corundum type $Al_2O_3$ formation [24,25].

Therefore, this study focuses on the thermal stability and oxidation resistance of arc evaporated $Ti_{1-x-y}Al_xTa_yN$ protective coatings and especially on the influence of Al and Ta concentrations on the oxide scale formation.

## 2. Experimental

$Ti_{1-x-y}Al_xTa_yN$ coatings (thickness ~3 µm) with six different compositions were reactively deposited in an industrial scale Balzers INNOVA arc evaporation plant using a target current of 150 A, a bias potential of -80 V, a substrate temperature of 500 °C and a nitrogen pressure of 3.5 Pa. The Prior to the depositions on various substrates (alumina (99.7% $Al_2O_3$), silicon (100), iron foil, austenite, and high speed steel), an argon ion etching step was used to further clean the substrate surfaces allowing for increased coating adherence. The depositions were conducted with $Ti_{0.50}Al_{0.50}$, $Ti_{0.475}Al_{0.475}Ta_{0.05}$, $Ti_{0.45}Al_{0.45}Ta_{0.10}$, $Ti_{0.34}Al_{0.66}$, $Ti_{0.323}Al_{0.627}Ta_{0.05}$, or $Ti_{0.306}Al_{0.594}Ta_{0.10}$ targets, which were powder metallurgically formed from single element Al, Ti, Ta metal powders (final density higher than 99 %).

Structural investigations were performed with a PANalytical X'pert X-Ray diffractometer (XRD) in Bragg Brentano geometry with CuKα radiation. Cross-sectional micrographs were obtained by a FEI Quanta 200 field emission gun (FEG) scanning electron microscope (SEM). The attached EDAX Pegasus XM4 energy dispersive X-ray (EDX) detector was used for evaluating the chemical composition. Differential scanning calorimetry (DSC) combined with thermo-gravimetric analysis (TGA) measurements were performed in a SETSYS Evolution 2400 DSC using a heating rate of 20 K/min, 50 sccm He flow, and a maximum temperature of 1600 °C. Isothermal DSC and TGA oxidation investigations were performed at 850 and 950 °C. Therefore, the samples were heated with 50 K/min in Helium and kept for 20 min at the respective temperature. Subsequently, the oxidation step was initiated by exchanging the He atmosphere with synthetic air (20.5 vol% $O_2$, 79.5 vol% $N_2$). After 5 h of oxidation, the samples were cooled down to room temperature with a cooling rate of 50 K/min. To investigate the oxide phase evolution with progressing oxidation time additional DSC-TGA oxidation measurements were performed for selected samples and specific treatment durations. To avoid substrate interference during DSC-TGA measurements the coatings were chemically removed (diluted nitric acid) from their iron foils. Prior and after these DSC-TGA measurements the coating materials were analyzed by XRD. Additionally to the TGA signal during oxidation, we have measured the weight of our samples before and after the oxidation treatment by an external high precision balance. This data was used for calculating the fraction of consumed nitride material (see chapter 3.2, Table 1).

Ambient air oxidation treatments of coated alumina samples were performed at 850 and 950 °C for 20 h. SEM fracture cross sectional micrographs of these samples were used for discussing the morphology of the respective oxide scale.

## 3. Results

The chemical compositions of our arc-evaporated coatings, normalized to a nitrogen content of 50 %, are $Ti_{0.51}Al_{0.49}N$, $Ti_{0.47}Al_{0.45}Ta_{0.08}N$, $Ti_{0.43}Al_{0.42}Ta_{0.15}N$, $Ti_{0.35}Al_{0.65}N$, $Ti_{0.32}Al_{0.60}Ta_{0.08}N$ and $Ti_{0.30}Al_{0.54}Ta_{0.16}N$. Two general trends can be observed when relating the coating compositions to the used target compositions. The Al content is always below and the Ta content significantly above (~50 %) the target composition. This can be explained by the high mass-difference between Al and Ta and hence Al could be significantly more scattered in the gas phase or resputtered from the growing film. Figure 1 shows fracture cross sectional micrographs of reactively arc evaporated Al rich (Al/(Al+Ti) ~0.65) $Ti_{1-x-y}Al_xTa_yN$ coatings on high speed steel with Ta contents of y = 0, 0.08, and 0.16, Fig. 1a-c. These images are representative for all investigated coatings and exhibit a dense and fine columnar structure. The corresponding XRD patterns, Figs. 2a and b, for medium and high Al contents, respectively, indicate a single phase face centered cubic B1 structure for all Ta contents. The crystallite size (evaluated with a quadratic Williamson Hall plot) is around 20 nm and slightly decreases with increasing tantalum content. According to a linear interpolation between the lattice parameters of fcc-TiN and fcc-AlN, as suggested by Vegard for highly ionic crystals [26] and hence not for such materials, the XRD peak positions for fcc-Ti$_{0.51}$Al$_{0.49}$N should be in the middle between fcc-TiN and fcc-AlN. Deviations from these XRD positions are due to the positive deviations from Vegard's estimation as studied by Density Functional Theory (DFT) in Refs. [10,27] and due to the physical vapor deposition process generated structural defects leading to micro-strains. With increasing tantalum content (Ti$_{0.47}$Al$_{0.45}$Ta$_{0.08}$N and Ti$_{0.43}$Al$_{0.42}$Ta$_{0.15}$N) clear peak shifts to lower diffraction angles (towards fcc-TaN) suggest for a solid solution of Ta to the metal sublattice. A corresponding behavior can be observed for our Al rich coatings with compositions of Ti$_{0.35}$Al$_{0.65}$N, Ti$_{0.32}$Al$_{0.60}$Ta$_{0.08}$N, and Ti$_{0.30}$Al$_{0.54}$Ta$_{0.16}$N, Fig. 2b.

### 3.1. Thermal stability

Differential scanning calorimetry of our samples clearly exhibit pronounced exothermic features in the temperature range 500 – 1300 °C, see Fig. 3. The broad exothermic feature between 500 and 1000 °C is the sum of several exothermic reactions like recovery effects (annihilation or rearrangement of defects to lower energy sites) and especially the decomposition of the supersaturated fcc-Ti$_{1-x-y}$Al$_x$Ta$_y$N phase to isostructural Ti rich and Al rich cubic domains, see Refs. [28,29] for more details. The main exothermic features with peak temperatures between 1000 and 1200 °C can be assigned to the phase transformation of the fcc Al rich cubic domains (or even fcc-AlN) into hcp-AlN. The massive volume expansion of ~24 % of this transformation [7,30] can lead to a disintegration of the coatings and increased crack network. Most likely this is also the reason for the often-observed connected mass-reduction, due to $N_2$-release driven by available diffusion pathways. The peak-temperatures of the described DSC feature increases from 1100 to 1200 to 1250 °C for the medium Al containing coatings with increasing Ta content. Correspondingly, also their mass-loss is shifted to higher temperatures from ~1200 to 1300 °C, see Fig. 3a. Whereas TGA suggest that the mass-loss of $Ti_{0.51}Al_{0.49}N$ is finished after a loss of ~2 %, the Ta containing coatings $Ti_{0.47}Al_{0.45}Ta_{0.08}N$ and $Ti_{0.43}Al_{0.42}Ta_{0.15}N$ still exhibit a small mass-reduction after the distinct mass-loss between 1300 – 1400 °C.

The decreasing area under the DSC curves is indicative for a decreased stored energy within the coatings with increasing Ta content. This is predicted by Holec et al. [11] as alloying $Ti_{1-x}Al_xN$ with tantalum should result in reduced mixing enthalpies and thereby in a reduced driving force for decomposition of the supersaturated phases into isostructural components. Consequently, the onset and intensity of the decomposition is shifted to higher temperatures and thereby also the subsequent phase transformation of Al rich cubic domains (or even fcc-AlN) to the stable modification hcp-AlN, see Fig. 3a. A corresponding behavior is obtained for our Al rich coatings. In sound agreement to previous studies on $Ti_{1-x}Al_xN$ coatings (sputtered or arc-evaporated) [27,28,31,32] the peak temperature of the pronounced exothermic feature (due to the phase transformation of fcc-AlN to hcp-AlN) shifts to lower temperatures with increasing Al content, compare Figs. 3a and b. This is based on the increased driving force for decomposition of the supersaturated $Ti_{1-x}Al_xN$ matrix into their components fcc-TiN and fcc-AlN with increasing Al content. As the peak-temperature of the fcc-hcp phase transformation of $Ti_{0.35}Al_{0.65}N$ with 1050 °C is at lower temperatures, as for the lower Al containing $Ti_{0.51}Al_{0.49}N$, also the mass-loss starts earlier, see Fig. 3b. Nevertheless, almost at the same temperature of 1300 °C the mass-loss is completed. Again the addition of Ta to $Ti_{1-x}Al_xN$ shifts the peak temperature for the fcc-hcp phase transformation to higher temperatures and hence also the mass-loss. However, the TGA curve suggests that with increasing Ta content the mass-loss still proceeds (with a significantly reduced rate) after the distinct mass-loss of ~1.3 and 2 % for the high and medium Al containing coatings, respectively. The decreasing mass-loss between high and medium Al containing coatings corresponds with a decreasing Ti content. This suggests that the nitrogen loss strongly correlates with the Ti content. According to the Ti-N phase diagram [33] fcc-TiN exhibits a broad chemical stability range and allows for a nitrogen release while still keeping the fcc structure. On the other hand AlN is known to be a line compound with only a very limited chemical stability range [34].

Powder XRD patterns after DSC-TGA measurements to 1600 °C clearly show that the Ta free coatings $Ti_{0.51}Al_{0.49}N$, Fig. 4a, and $Ti_{0.35}Al_{0.65}N$, Fig. 4b, are completely decomposed into their stable constituents fcc-TiN and hcp-AlN. With increasing Ta content the positions of the fcc XRD peaks shift to lower diffraction angles from fcc-TiN to fcc-TaN, indicative for the solid solution between TiN and TaN. For the high Ta containing coatings $Ti_{0.43}Al_{0.42}Ta_{0.15}N$ and $Ti_{0.30}Al_{0.54}Ta_{0.16}N$ the formation of hexagonal (hex) $Ta_2N$ [35] can be detected by XRD. This is especially pronounced for the high Al containing

Ti$_{0.30}$Al$_{0.54}$Ta$_{0.16}$N, compare Figs. 4a and b, and in agreement to TGA measurements, Fig. 3, which exhibit an increasing mass-loss with increasing Ta content.

3.2. Oxidation Resistance

At the initial stage of oxidation the mass-gain $w$ usually follows a linear rate law with time $t$:

$$w = k_l t \qquad (1)$$

where $k_l$ the linear rate constant. The rate limiting reaction step of this stage is the surface reaction of oxygen or the diffusion through the gas phase and is often denoted as "Regime 1" [36]. As soon as a dense oxide scale is formed on top of the nitride coating the mass-gain rate is retarded and controlled by the diffusion of the reactants through the scale. This bulk/grain-boundary diffusion is based on Fick's diffusion law and here the mass-gain usually follows a parabolic rate law ("Regime 2"):

$$w = \sqrt{k_p t} \qquad (2)$$

where $k_p$ is the parabolic rate constant [37,38]. In the case of high temperature oxidation, often a paralinear behavior is observed, where a predominant parabolic Regime 2 is accompanied by a linear rate-law:

$$w = \sqrt{k_p t} + c k_l t \qquad (3)$$

with a reduced (by the prefactor c; usually much smaller than 1) linear rate constant. This is the case if the formed oxide scale suffers damage e.g. due to cracks [39]. As soon as the parabolic Regime 2 is reached, the mass-gain and hence film degradation can effectively be retarded and hence $k_p$ is a characteristic parameter for describing the quality of the formed scale. $k_p$ is usually given as

$$k_p = \left(\frac{\Delta m}{A}\right)^2 \cdot \frac{1}{t} \qquad (4)$$

, where the mass gain $\Delta m$ is related to the exposed surface area $A$ and time $t$. Within this study we have used flake-like coating material, which can be described by plates having a top and a bottom surface area $A$ and a thickness d (equal to the coating thickness d). If these plates are large (their volume to surface ratio high) their lateral surface area (coating thickness times plate perimeter) can be neglected and the rate constant for flake-like coating powder $k_p^*$ can be related to $k_p$ by

$$k_p^* \cdot \left(\frac{\rho d}{2}\right)^2 = k_p \qquad (5)$$

, where $\rho$ is the density of the coating. Consequently the rate constant is defined as

$$k_p^* = \left(\frac{\Delta m}{m_0}\right)^2 \cdot \frac{1}{t} \qquad (6)$$

, where $m_0$ is the weighed portion of the flake-like nitride coating material. The factor $\left(\frac{\rho d}{2}\right)^2$ is different for the different coatings, but does not influence activation energy evaluations by Arrhenius plots as it is constant for the different treatment temperatures.

Figure 5a shows the isothermal DSC and TGA curves of $Ti_{0.51}Al_{0.49}N$, $Ti_{0.47}Al_{0.45}Ta_{0.08}N$, and $Ti_{0.43}Al_{0.42}Ta_{0.15}N$ in synthetic air at 850 °C. The DSC signal clearly shows that the exothermic signal initially rapidly increases and soon decreases again. The increase in heat-flow corresponds to the formation of an oxide scale on top of the nitride layer. A decreasing heat flow after the peak value suggests either a completed nitride-to-oxide transformation or that the oxidation process is effectively retarded by the formation of a dense protective oxide. The latter is the case for our coatings as the TGA signal shows a pronounced reduction in mass-gain-rate without reaching the maximum mass-gain due to a completed nitride-to-oxide transformation. After a linear Regime 1, also in our case the Ta free $Ti_{0.51}Al_{0.49}N$ coating follows a parabolic growth rate at 850 °C with $k_p^*$ = 1.83 x $10^{-3}$ 1/s (parabolic fits are indicated by the dashed lines). Again increasing mass-gain rates after 60 min of oxidation suggest for a failure (e.g. cracks) of the formed oxide scale. The small inset in Fig. 5a shows a magnification of the first 9 minutes of the DSC signal, indicating that the onset of the exothermic signal due to oxidation is similar for all three coatings. Nevertheless, the intensity as well as the time of the oxidation peak decreases with increasing Ta content. Even more important is that the exothermic output after this peak-reaction also decreases with increasing Ta content. This clearly suggests that with increasing Ta content the oxide scale formed is denser and hence more effective in reducing the oxidation process. This is proven by the TGA measurements showing that

Ti$_{0.51}$Al$_{0.49}$N is already fully oxidized after ~120 min as the mass-gain levels out at 22 %. This value is close to the theoretical possible mass-gain of ~27 % if the nitride fully transforms into TiO$_2$ and Al$_2$O$_3$. Deviations from the theoretical mass-gain can be explained by the commonly observed over-stoichiometry of Ti$_{1-x}$Al$_x$N and the sub-stoichiometry of TiO$_2$ [1,19,40]. As a consequence, this leads to a higher mass-loss due to the nitrogen release (from overstoichiometric nitrides as compared to stoichiometric nitrides) and a lower mass-gain due to oxidation (to substoichiometric TiO$_2$ as compared to stoichiometric TiO$_2$). At that point we also want to notice that an EDX chemical analysis is especially for light elements (like N) not very precise and therefore this might have also an influence on the measured mass-gain during oxidation.

The mass-gain-rate is effectively reduced with increasing Ta content and follows a parabolic-like growth with constants of $k_p^*$ = 7.2 × 10$^{-4}$ and 5.8 × 10$^{-4}$ 1/s for Ti$_{0.47}$Al$_{0.45}$Ta$_{0.08}$N and Ti$_{0.43}$Al$_{0.42}$Ta$_{0.15}$N, respectively. Even after 5 h exposure to synthetic air at 850 °C the coatings are far from being fully oxidized.

The Al rich coatings show an even improved oxidation resistance, Fig. 5b. The initial oxidation process within the first minutes of exposure to synthetic air is significantly reduced and the heat flow decreases earlier. Compare the insets in Figs. 5a and b. This suggests that the oxide scale formed is even denser as for their lower Al containing counterparts, which is also confirmed by the mass-gain-curves. Within the first two hours the mass-gain-curve of Ti$_{0.35}$Al$_{0.65}$N can also be described by a first linear Regime 1 which is followed by a parabolic behavior with $k_p^*$ = 3.15 × 10$^{-4}$ 1/s, to be followed by a linear-like oxidation behavior. The latter is indicative for the formation of many cracks within the scale ~~or even spallation of parts of the oxide scale~~ and hence increased diffusion and exposure of unprotected – by the oxide scale – nitride material. Nevertheless, the mass-gain of ~12 % after 5 h indicates that only ~1/2 of the nitride coating is transferred into TiO$_2$ and Al$_2$O$_3$. For a completed oxidation of Ti$_{0.35}$Al$_{0.65}$N the theoretical mass-gain would be ~26.5 %. Corresponding to the medium Al containing coatings, the oxidation resistance can effectively be increased with increasing Ta content. The parabolic growth rate Regime 2 of the mass-gain-curves of Ti$_{0.32}$Al$_{0.60}$Ta$_{0.08}$N and Ti$_{0.30}$Al$_{0.54}$Ta$_{0.16}$N can be described with $k_p^*$ = 1.23 × 10$^{-4}$ and $k_p^*$ = 1.60 × 10$^{-4}$ 1/s. Increasing the oxidation temperature to 950 °C is extremely demanding for Ta free Ti$_{0.51}$Al$_{0.49}$N and Ti$_{0.35}$Al$_{0.65}$N. For both coatings, DSC can detect a pronounced exothermic output within the first 15 min of exposure, and the mass-gain also suggests that the coatings are soon completely oxidized, see Figs. 6a and b. Their mass-gain follows a linear-like behavior. The higher Al containing coating slows down the initially pronounced oxidation process within the first five minutes, as suggested by the decreasing heat flow and the slightly reduced mass-gain rate. Therefore, we were also able to determine the parabolic rate-constant for this short term with $k_p^*$ = 2.40 × 10$^{-3}$ 1/s. But already after ~2 min, again enhanced oxidation reactions occur leading to a pronounced linear-like increase in mass. ~~This behavior suggests a pronounced breakaway oxidation process.~~ The lower Al containing coating, Ti$_{0.51}$Al$_{0.49}$N, exhibits no formation of a dense oxide, which is indicated by only linear growth in the mass gain curve from the beginning to the end. The heat-flow only decreases at the end of the oxidation process (~12 min) when the coating is completely oxidized, suggesting that here not even the initial stage of a protecting oxide scale (Regime 2) can be formed.

However, for both coating types (medium and high Al content) the addition of Ta significantly improves their oxidation resistance, as the exothermic output as well as the mass-gain is significantly reduced. For the medium Al containing films the heat flow decreases after ~3 and 2 min for $Ti_{0.47}Al_{0.45}Ta_{0.08}N$ and $Ti_{0.43}Al_{0.42}Ta_{0.15}N$, respectively, indicating the formation of a protective oxide scale. The mass-gain of $Ti_{0.43}Al_{0.42}Ta_{0.15}N$ follows a parabolic growth rate with $k_p^*$ = 4.50 × $10^{-4}$ 1/s until the maximum testing time of 5 h, whereas for the lower Ta containing $Ti_{0.47}Al_{0.45}Ta_{0.08}N$ only the first hour can be described by a parabolic growth rate with $k_p^*$ = 1.13 × $10^{-3}$ 1/s. The latter exhibits after ~1 h an increased mass-gain, corresponding to paralinear oxidation (see equation 3) due to significant cracking of the protective oxide scale, which results after 5 h to a completed oxidation. Contrary, the higher Ta containing coating does not show such an additional linear oxidation behavior and the mass-gain suggests that only ~57 % of the nitride is transferred to $TiO_2$, $Al_2O_3$, and $Ta_2O_5$ after 5 h, Fig. 6a. With increasing Al content the Ta addition is even more effective, but of utmost importance is, that here the lower Ta containing $Ti_{0.32}Al_{0.60}Ta_{0.08}N$ follows the parabolic mass-gain with a smaller rate constant $k_p^*$ = 1.40 x $10^{-4}$ 1/s than that of the higher Ta containing counterpart $Ti_{0.30}Al_{0.54}Ta_{0.16}N$ ($k_p^*$

= 2.40 × 10$^{-4}$ 1/s), Fig. 6b. The mass-gain (measured by an external balance; see Table 1) of 6.4 and 10.2% after 5 h exposure to air at 950 °C suggests that 28 and 49% of Ti$_{0.32}$Al$_{0.60}$Ta$_{0.08}$N and Ti$_{0.30}$Al$_{0.54}$Ta$_{0.16}$N are oxidized, respectively.

XRD investigations of the Ta free coatings Ti$_{0.51}$Al$_{0.49}$N and Ti$_{0.35}$Al$_{0.65}$N after isothermal DSC-TGA in air for 5 h at 850 °C reveal a rutile (r-TiO$_2$) [41], anatase (a-TiO$_2$) [42], and corundum ($\alpha$-Al$_2$O$_3$) [43] type oxide scale structure, Figs. 7a and b. For the higher Al containing Ti$_{0.35}$Al$_{0.65}$N coating, additional broad XRD peaks at diffraction angles of ~37° and ~43° suggest the presence of fcc-Ti$_{1-x}$Al$_x$N. Furthermore, also indications for monoclinic Al oxides (m-AlO$_x$ [44]) and hcp-AlN can be detected. This is in agreement to TGA suggesting no completed oxidation of the nitride layer during the 5 h oxidation treatment. Here, the fcc-AlN to hcp-AlN transformation could be additionally responsible for the again increasing mass-gain-rate at an exposure time of 180 min. With the addition of Ta to both, medium and high Al containing Ti$_{1-x}$Al$_x$N, coatings the formation of the anatase phase is suppressed and the oxide scale predominantly exhibits rutile. This does not necessarily imply that Al has to be dissolved in a rutile phase, as especially an Al oxide – if very thin – is difficult to detect and can be amorphous like [8,24].

While the intensity of the rutile phase decreases with increasing Ta content, the overall contribution of the Ti$_{1-x-y}$Al$_x$Ta$_y$N phase increases. Furthermore, no hcp-AlN can be detected for these Ta containing nitrides. These results are in excellent agreement to DSC-TGA in inert (He) and oxidizing atmosphere (synthetic air) showing that with the addition of Ta the hcp-AlN formation is shifted to higher temperatures and only a small fraction of the nitride coating is oxidized. Increasing the oxidation temperature to 950 °C leads also – in addition to the full oxidation of $Ti_{0.51}Al_{0.49}N$ Fig. 8a – to a complete oxidation of the Al rich $Ti_{0.35}Al_{0.65}N$, see Fig. 8b. There are no detectable remaining nitride phases present for these coatings after the oxidation at 950 °C for 5 h. The oxides are composed of rutile and corundum type phases, no anatase can be detected by XRD. For this high temperature, also the Ta containing $Ti_{0.47}Al_{0.45}Ta_{0.08}N$ (lowest Ta and medium Al content) is completely oxidized, Fig. 8a, as already suggested by the corresponding DSC-TGA curves, see Fig. 6a. A further increase in Ta content to form $Ti_{0.43}Al_{0.42}Ta_{0.15}N$ allows for the formation of a protective dense oxide scale and hence, the presence of fcc-$Ti_{1-x-y}Al_xTa_yN$ can still be detected (broad XRD peaks at 2θ angles of ~37° and 43°). For higher Al containing coatings this is valid even for the lower Ta content, see Fig. 8b. Similar to the structure after oxidation at 850 °C the oxides are predominantly composed of rutile phase (no anatase), but due to the higher temperature of 950 °C, here also low intensity α-$Al_2O_3$ can be detected.

With the help of Arrhenius plots, we calculate activation energies for the formation of a dense oxide scale using the parabolic rate constants $k_p^*$ (Regime 2). This yields ~232 kJ/mol for the high Al containing coating $Ti_{0.35}Al_{0.65}N$. Tantalum additions lead to significantly reduced activation energies of 14.8 kJ/mol for $Ti_{0.32}Al_{0.60}Ta_{0.08}N$ and 46.3 kJ/mol for $Ti_{0.30}Al_{0.54}Ta_{0.16}N$. Consequently, the energy barrier for the stable Regime 2 behavior is reduced by a factor of ~5 for $Ti_{0.30}Al_{0.54}Ta_{0.16}N$ and ~15 for $Ti_{0.32}Al_{0.60}Ta_{0.08}N$ as compared to conventional $Ti_{0.35}Al_{0.65}N$, indicating that the passivation process is fastest and easiest for $Ti_{0.32}Al_{0.60}Ta_{0.08}N$. For medium Al containing $Ti_{0.47}Al_{0.45}Ta_{0.08}N$, the activation energy with 51.5 kJ/mol is comparable to $Ti_{0.30}Al_{0.54}Ta_{0.16}N$. Please note that we have used $k_p^*$ of only two different temperatures (850 and 950 °C) and hence these activation energies can just be used for rating the different coatings within this study.

Table 1 summarizes the above presented data for our coatings, obtained by XRD and isothermal DSC-TGA at 850 and 950 °C. This allows for a clear and detailed comparison of the individual effect of Al and Ta on the oxidation behavior of our $Ti_{1-x-y}Al_xTa_yN$ coatings.

Generally, the two following major conclusions can be drawn from the oxidation experiments. As long as the nitride coating exhibits a single phase structure and no nitride phase transformations occur at the oxidation temperature, increasing Al contents significantly reduce the oxidation processes. The addition of Ta to form single-phase fcc-$Ti_{1-x-y}Al_xTa_yN$ further reduces the oxidation process by promoting the formation of a protective dense oxide scale. However, the detailed investigations clearly show that the Ta content needs to be balanced with the Al content. Moreover, the ideal chemical composition for optimized oxidation resistance also depends on the application temperature.

Cross sectional SEM investigations of the coatings after exposure to ambient air for 20 h at 850 and 950 °C clearly show that the Ta free coatings $Ti_{0.51}Al_{0.49}N$ and $Ti_{0.35}Al_{0.65}N$ are fully oxidized and exhibit a porous morphology, like shown in Fig. 9a for $Ti_{0.35}Al_{0.65}N$. As suggested by DSC-TGA the Ta alloyed coatings are able to form a dense oxide scale during exposure to air at 850 °C, allowing for a reduced oxidation rate and a parabolic-like growth behavior. Such a dense oxide scale formation, after 20 h at 850 °C, is shown for $Ti_{0.43}Al_{0.42}Ta_{0.15}N$ in Fig. 9b. However, by increasing the temperature to 950 °C the oxides formed on $Ti_{0.47}Al_{0.45}Ta_{0.08}N$, $Ti_{0.43}Al_{0.42}Ta_{0.15}N$, as well as $Ti_{0.30}Al_{0.54}Ta_{0.16}N$ are composed of a porous region at the coating-scale interface and a denser outermost region. This is shown in Fig. 9c for $Ti_{0.30}Al_{0.54}Ta_{0.16}N$, which performed best during DSC-TGA at 950 °C among these three Ta containing coatings. The coating with the smallest parabolic growth rate constant ($k_p^*$ of 1.40 x $10^{-4}$ 1/s) during isothermal oxidation at 950 °C is $Ti_{0.32}Al_{0.60}Ta_{0.08}N$ (see table 1), which exhibits only a very thin (~100 nm) porous region at the interface between nitride and the dense outermost oxide scale, see Fig. 9d. Please notice that the magnification for Fig. 9d is increased as compared to Figs. 9a-c to clearly present the only ~1 µm thin oxide scale.

The presented results on the oxidation behavior of as-deposited single-phase cubic $Ti_{1-x-y}Al_xTa_yN$ with the main data summarized in Table 1 raise the question on the role of the anatase phase, its dependence on the Ta content, and the role of the chemical composition, which will be discussed in the <span style="color:red">next section</span>.

**Discussion**

Oxidizing the Ta free $Ti_{0.51}Al_{0.49}N$ and $Ti_{0.35}Al_{0.65}N$ at 850 °C causes the formation of rutile, anatase, and corundum phase fractions. By increasing the temperature to 950 °C no anatase can be detected anymore, and the oxides consist of only rutile and corundum phases. For a better understanding of the mechanisms behind, we have stopped the isothermal DSC-TGA treatments after specific times of oxidation. These samples are analyzed in detail by XRD. If $Ti_{0.51}Al_{0.49}N$ is oxidized for 2.5 min at 850 °C (directly after the first exothermic peak, see Fig. 5a) mainly anatase, only small amounts of rutile, and almost no corundum phases can be detected, Fig. 10a. After this short exposure to air the nitride phase fcc-$Ti_{1-x}Al_xN$ is still clearly present, please compare the corresponding XRD pattern with the as-deposited pattern at the bottom. Increasing the exposure time to 120 min (where the mass-gain reached its saturation, Fig. 5a) results in a completed oxidation of the nitride into anatase, rutile, and corundum phases. Upon further increasing the exposure time to 300 min (5 h) the rutile phase fraction increases at the expense of anatase. Generally, the anatase-to-rutile phase transformation would occur in the temperature range 550 – 700 °C [45,46] but the presence of other alloying elements like Al or N can further influence this transformation temperature [47,48]. However, also for the higher Al containing coating $Ti_{0.35}Al_{0.65}N$ the formation of anatase can be detected at an even higher oxidation temperature of 950 °C during the initial stage of oxidation. The XRD pattern taken after 4 min of exposure at 950 °C, Fig. 10b, clearly shows the presence of anatase, small fractions of rutile, and corundum phases in addition to the nitride phases hcp-AlN, fcc-Ti$_{1-x}$Al$_x$N, and fcc structured Ti and Al rich domains. The latter originate from spinodal decomposition of the supersaturated Ti$_{1-x}$Al$_x$N matrix [6,29,31,49]. Upon further exposure no nitrides can be detected anymore and the oxide phase anatase completely transforms into rutile. Based on these detailed XRD investigations we can conclude that the initial Ti oxide formed on medium and high Al containing single-phase fcc-Ti$_{1-x}$Al$_x$N is anatase, at least in this industrially important temperature range up to 950 °C. Due to the high temperature, anatase continuously transforms into rutile with progressing oxidation time. As this transformation is accompanied by a volume shrinkage of 5-10 % (rutile has a lower specific volume than anatase) [50,51], the formation of cracks and/or pores (Fig. 9a) is promoted leading to paralinear or even linear mass-gain-rates, Figs. 5 and 6. The formation of pores, especially at interfaces, also originates from different diffusion rates of the involved species, which is known as Kirkendall effect [52,53]. Especially at temperatures above 800 °C also the rapid outward diffusion of Al easily leads to the formation of voids [1,9,54].

The oxidation behavior of Ta alloyed Ti$_{1-x}$Al$_x$N coatings is especially different with respect to the anatase formation, which in contrast to Ti$_{1-x}$Al$_x$N cannot be observed by XRD during oxidation at 850 or 950 °C. Small fractions of crystalline corundum oxide can only be detected for the treatment at 950 °C, see Figs. 8a and b. This is in agreement to the oxide scale morphology of Ta alloyed coatings being uniform in contrast to Ti$_{1-x}$Al$_x$N, Fig. 9. As rutile TaO$_2$ is one of the most stable substoichiometric phases of Ta$_2$O$_5$ [55,56], we suggest that Ta stabilizes the formation of rutile phases already during the initial stages of oxidation of Ta alloyed $Ti_{1-x}Al_xN$. Based on the XRD experiments, exhibiting almost a perfect match of the rutile oxide peaks to the diffraction file of r-$TiO_2$, we envision that pentavalent Ta enables also the solubility of trivalent aluminum in rutile. This would also fit to the observation that the intensity of the $\alpha$-$Al_2O_3$ XRD peaks decreases with increasing Ta content, as the higher Ta content within the rutile phase allows also for a higher solubility of Al. The increasing broadening of the corresponding XRD peaks further supports this argument, as this indicates higher micro-strains and/or smaller grain sizes. Excess aluminum, which is not solved in the rutile phase, is responsible for the formation of $\alpha$-$Al_2O_3$, especially at high temperatures of 950 °C, as $\alpha$-$Al_2O_3$ is the stable oxide [57].

In spite of this, a major advantage of alloying Ta to $Ti_{1-x}Al_xN$ is the suppression of the anatase phase formation and the promotion of the high temperature rutile phase formation upon exposure to air. This is also supported by *ab initio* calculations and hence the destructive anatase-rutile phase transformation, which is typical during oxidation of $Ti_{1-x}Al_xN$ (as shown here and in [58] for example), is avoided. As mentioned above, the anatase-rutile phase transformation is connected with a contraction of 5-10 % [50,51] leading to crack formation and promotion of pores. We suggest, that the suppression of phase transformations within oxide scales is at least as important as reducing the bulk diffusion (as mentioned in earlier studies [24]), to reduce the oxide scale growth rate. ~~This is especially valid for polycrystalline oxide scales on various materials, where the usually faster grain boundary diffusion is the rate-controlling factor.~~

## Summary and Conclusion

We have used DSC, TGA, SEM, and XRD to study in detail the thermal stability and oxidation resistance of arc-evaporated $Ti_{1-x-y}Al_xTa_yN$ hard coatings. By adding Ta up to y = 0.15 we could show that the stored mixing energy is significantly decreased and hence the thermal decomposition into the stable compounds is increased by ~200 °C. Considering the oxide scale formation of $Ti_{1-x}Al_xN$ during oxidation at 850°C and 950 °C, the anatase-to-rutile phase transformation, which is accompanied by a volume change of 5 – 10 %, leads to the formation of pores, cracking, and maybe also to spallation of the scale. Consequently, only high Al containing $Ti_{0.65}Al_{0.35}N$ could resist complete oxidation after 5 h at 850 °C, whereas at 950 °C medium and high Al containing coatings were oxidized within 15 min mainly due to a linear mass-gain behavior. Tantalum additions lead in all considered cases to an early formation of rutile phases with reduced or even completely suppressed anatase formation. Based on these results we conclude that pentavalent Ta enables the solubility of trivalent Al in a rutile structured $TiO_2$ and the anatase-to-rutile phase transformations are avoided. Especially $Ti_{0.32}Al_{0.60}Ta_{0.08}N$ exhibits an extremely dense oxide scale after oxidation at 950 °C in contrast to the other coatings. Consequently, here the Al and Ta content is well-balanced to guarantee for stable single phase nitride coating at the oxidation temperature and to allow for a direct and fast formation of rutile $(Ti,Al,Ta)O_2$ and corundum $Al_2O_3$.

These results strongly indicate that the chemical composition of X alloyed $Ti_{1-x}Al_xN$ coatings has to be optimized to the ternary $(Ti,Al,X)_xO_y$ phase diagram to avoid anatase phase formation and hence gain significantly enhanced oxidation resistance. This could widely open up the field of applications of Ti-Al-N based hard coatings.

## Acknowledgements

The financial support by the Austrian Federal Ministry of Economy, Family and Youth and the National Foundation for Research, Technology and Development is greatly acknowledged. The authors also want to thank Harald Leitner and Christopher Pöhl for their contribution to the DSC/TGA investigations. SEM investigations were carried out using facilities at the University Service Centre for Transmission Electron Microscopy (USTEM), Vienna University of Technology, Austria. Thanks are also due to the X-ray center (XRC) of Vienna University of Technology, Austria.

# Tables

Table 1: Consumed coating material (calculated from the experimental mass-gain with respect to the theoretical mass-gain due to a complete transformation of $Ti_{1-x-y}Al_xTa_yN$ into 1-x-y $TiO_2$, x $Al_2O_3$ and y $Ta_2O_5$), formed oxide phases (r: rutile, a: anatase, $\alpha$: corundum, m: monoclinic $AlO_x$) after 5 h of oxidation, oxidation rate constants $k_p^*$ for the parabolic regime (Regime 2), and activation energy $E_A$ for Regime 2 of $Ti_{1-x-y}Al_xTa_yN$ coatings.

| coating | consumed coating material | | oxide phases | | rate constant, $k_p^*$ $\left(\left(\frac{\Delta m}{m_0}\right)^2 * \frac{1}{t}\right)$ | | activation energy, $E_A$ |
|---|---|---|---|---|---|---|---|
| | 850 °C [%] | 950 °C [%] | 850 °C | 950 °C | 850 °C [1/s] | 950 °C [1/s] | [kJ/mol] |
| $Ti_{0.51}Al_{0.49}N$ | 83.3[a] | 83.5[a] | r+a+$\alpha$ | r+$\alpha$ | $1.83 * 10^{-3}$ | -[b] | - |
| $Ti_{0.47}Al_{0.45}Ta_{0.08}N$ | 43.8 | 87.2[a] | r+$\alpha$ | r+$\alpha$ | $7.20 * 10^{-4}$ | $1.13 * 10^{-3}$ | 51.5 |
| $Ti_{0.43}Al_{0.42}Ta_{0.15}N$ | 43.2 | 56.9 | r | r+$\alpha$ | $5.80 * 10^{-4}$ | $4.50 * 10^{-4}$ | -[c] |
| <span style="color:red">$Ti_{0.35}Al_{0.65}N$</span> | 51.7 | 89.5[a] | r+a+$\alpha$+m | r+$\alpha$ | $3.15 * 10^{-4}$ | $2.40 * 10^{-3}$ | 232 |
| $Ti_{0.32}Al_{0.60}Ta_{0.08}N$ | 26.5 | 27.8 | r | r+$\alpha$ | $1.23 * 10^{-4}$ | $1.40 * 10^{-4}$ | 14.8 |
| $Ti_{0.30}Al_{0.54}Ta_{0.16}N$ | 20.9 | 48.9 | r | r+$\alpha$ | $1.60 * 10^{-4}$ | $2.40 * 10^{-4}$ | 46.3 |

[a] these coatings are actually fully oxidized, the difference in mass-gain to the theoretical mass-gain is related to possible nitrogen over-stoichiometry of the nitride phase and oxygen under-stoichiometry of the formed oxide phases.

[b] just linear Regime 1 oxidation behavior detectable

[c] here the Arrhenius plot shows actually a positive slope

**Figure captions**

Fig.1: SEM fracture cross sections of Al rich coatings (a) $Ti_{0.35}Al_{0.65}N$, (b) $Ti_{0.32}Al_{0.60}Ta_{0.08}N$, and (c) $Ti_{0.30}Al_{0.54}Ta_{0.16}N$ in their as deposited state on high speed steel substrates.

Fig. 2: XRD patterns of (a) medium Al containing coatings $Ti_{0.51}Al_{0.49}N$, $Ti_{0.47}Al_{0.45}Ta_{0.08}N$, $Ti_{0.45}Al_{0.43}Ta_{0.15}N$, and (b) high Al containing coatings $Ti_{0.35}Al_{0.65}N$, $Ti_{0.32}Al_{0.60}Ta_{0.08}N$, and $Ti_{0.30}Al_{0.54}Ta_{0.16}N$ in their as deposited state (after removing from their Fe-foil substrates). The peak positions for fcc-TiN, fcc-TaN, and fcc-AlN are taken from the ICDD database [35,59,60].

Fig. 3: Dynamical DSC and TGA measurements of (a) medium Al containing coatings $Ti_{0.51}Al_{0.49}N$, $Ti_{0.47}Al_{0.45}Ta_{0.08}N$, $Ti_{0.45}Al_{0.43}Ta_{0.15}N$, and (b) high Al containing coatings $Ti_{0.35}Al_{0.65}N$, $Ti_{0.32}Al_{0.60}Ta_{0.08}N$, and $Ti_{0.30}Al_{0.54}Ta_{0.16}N$ in inert He atmosphere.

Fig. 4: XRD patterns of (a) medium Al containing coatings $Ti_{0.51}Al_{0.49}N$, $Ti_{0.47}Al_{0.45}Ta_{0.08}N$, $Ti_{0.45}Al_{0.43}Ta_{0.15}N$, and (b) high Al containing coatings $Ti_{0.35}Al_{0.65}N$, $Ti_{0.32}Al_{0.60}Ta_{0.08}N$, and $Ti_{0.30}Al_{0.54}Ta_{0.16}N$ after DSC-TGA to 1600 °C in inert He atmosphere, Fig. 3.

Fig. 5: Isothermal DSC and TGA measurements of (a) medium Al containing coatings $Ti_{0.51}Al_{0.49}N$, $Ti_{0.47}Al_{0.45}Ta_{0.08}N$, $Ti_{0.45}Al_{0.43}Ta_{0.15}N$, and (b) high Al containing coatings $Ti_{0.35}Al_{0.65}N$, $Ti_{0.32}Al_{0.60}Ta_{0.08}N$, and $Ti_{0.30}Al_{0.54}Ta_{0.16}N$ in synthetic air at 850 °C. Parabolic fits are indicated by the dashed lines.

Fig. 6: Isothermal DSC and TGA measurements of (a) medium Al containing coatings Ti$_{0.51}$Al$_{0.49}$N, Ti$_{0.47}$Al$_{0.45}$Ta$_{0.08}$N, Ti$_{0.45}$Al$_{0.43}$Ta$_{0.15}$N, and (b) high Al containing coatings Ti$_{0.35}$Al$_{0.65}$N, Ti$_{0.32}$Al$_{0.60}$Ta$_{0.08}$N, and Ti$_{0.30}$Al$_{0.54}$Ta$_{0.16}$N in synthetic air at 950 °C. Parabolic fits are indicated by the dashed lines.

Fig. 7: XRD patterns of (a) medium Al containing coatings Ti$_{0.51}$Al$_{0.49}$N, Ti$_{0.47}$Al$_{0.45}$Ta$_{0.08}$N, Ti$_{0.45}$Al$_{0.43}$Ta$_{0.15}$N, and (b) high Al containing coatings Ti$_{0.35}$Al$_{0.65}$N, Ti$_{0.32}$Al$_{0.60}$Ta$_{0.08}$N, and Ti$_{0.30}$Al$_{0.54}$Ta$_{0.16}$N after DSC-TGA in synthetic air at 850 °C.

Fig 8: XRD patterns of (a) medium Al containing coatings Ti$_{0.51}$Al$_{0.49}$N, Ti$_{0.47}$Al$_{0.45}$Ta$_{0.08}$N, Ti$_{0.45}$Al$_{0.43}$Ta$_{0.15}$N, and (b) high Al containing coatings Ti$_{0.35}$Al$_{0.65}$N, Ti$_{0.32}$Al$_{0.60}$Ta$_{0.08}$N, and Ti$_{0.30}$Al$_{0.54}$Ta$_{0.16}$N after DSC-TGA synthetic air at 950 °C.

Fig. 9: SEM fracture cross sections of representative coatings after 20 h exposure to ambient air at 850 °C (a, b) and 950 °C (c, d). (a) Fully oxidized Ti$_{0.35}$Al$_{0.65}$N. (b) Ti$_{0.43}$Al$_{0.42}$Ta$_{0.15}$N with a dense oxide scale on top. (c) Ti$_{0.30}$Al$_{0.54}$Ta$_{0.16}$N with a pronounced porous region between the coating and the dense oxide scale on top. (d) Ti$_{0.32}$Al$_{0.60}$Ta$_{0.08}$N with a strongly reduced porous region between coating and dense oxide scale.

Fig. 10: XRD patterns of the Ta free coatings after various durations of isothermal DSC-TGA in synthetic air at (a) 850 °C for Ti$_{0.51}$Al$_{0.49}$N and (b) 950 °C for Ti$_{0.35}$Al$_{0.65}$N. For comparison also their XRD patterns in the as deposited state are given.

# Figures

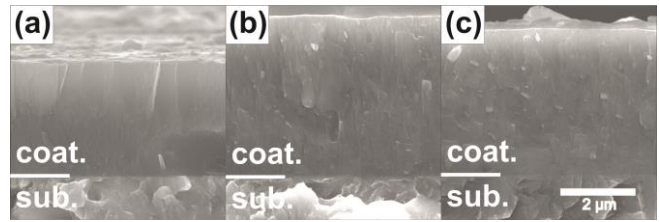

Fig.1

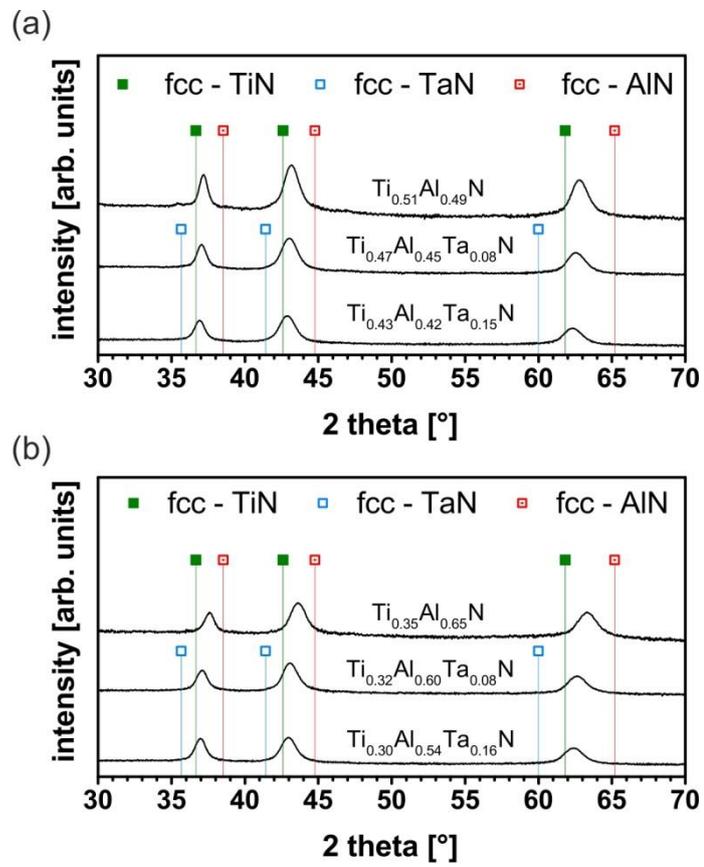

Fig.2

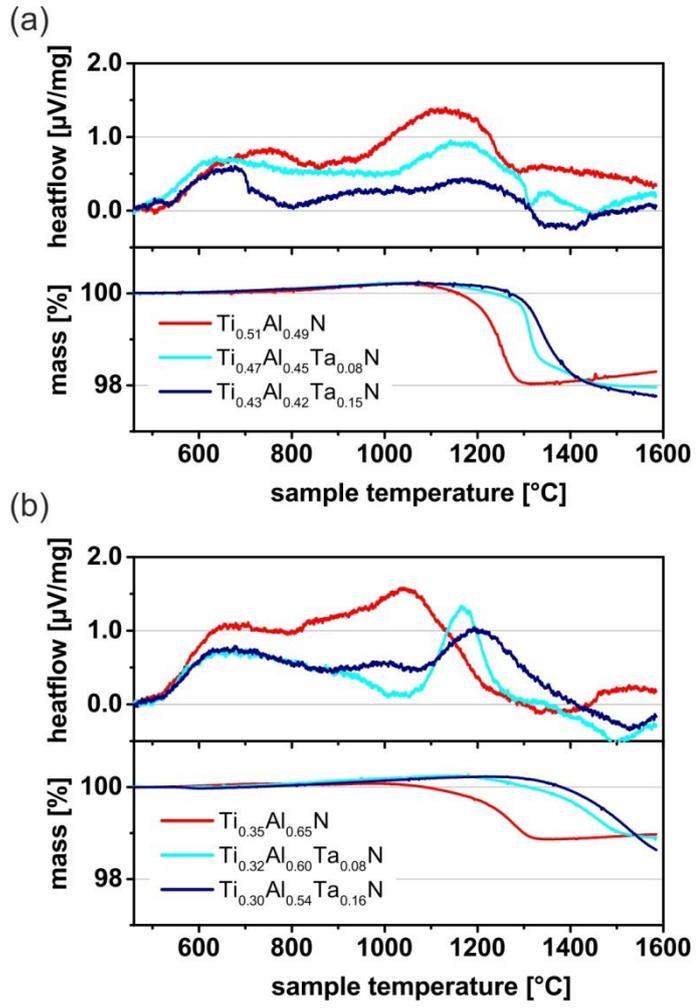

Fig.3

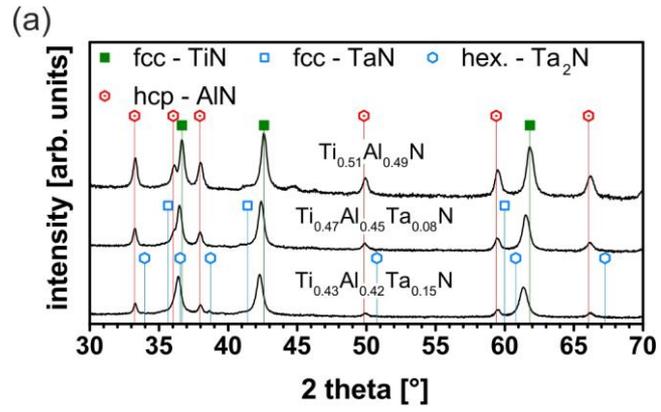

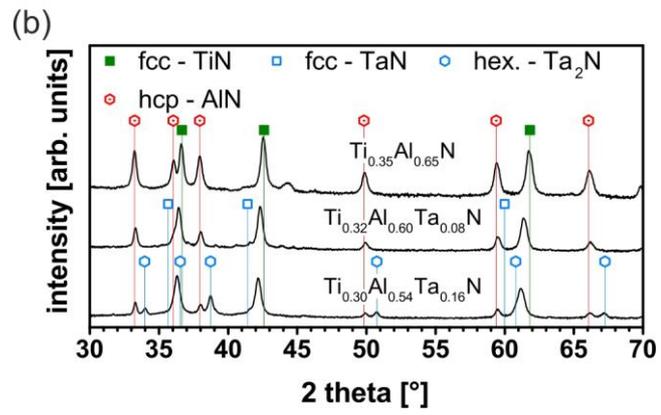

Fig.4

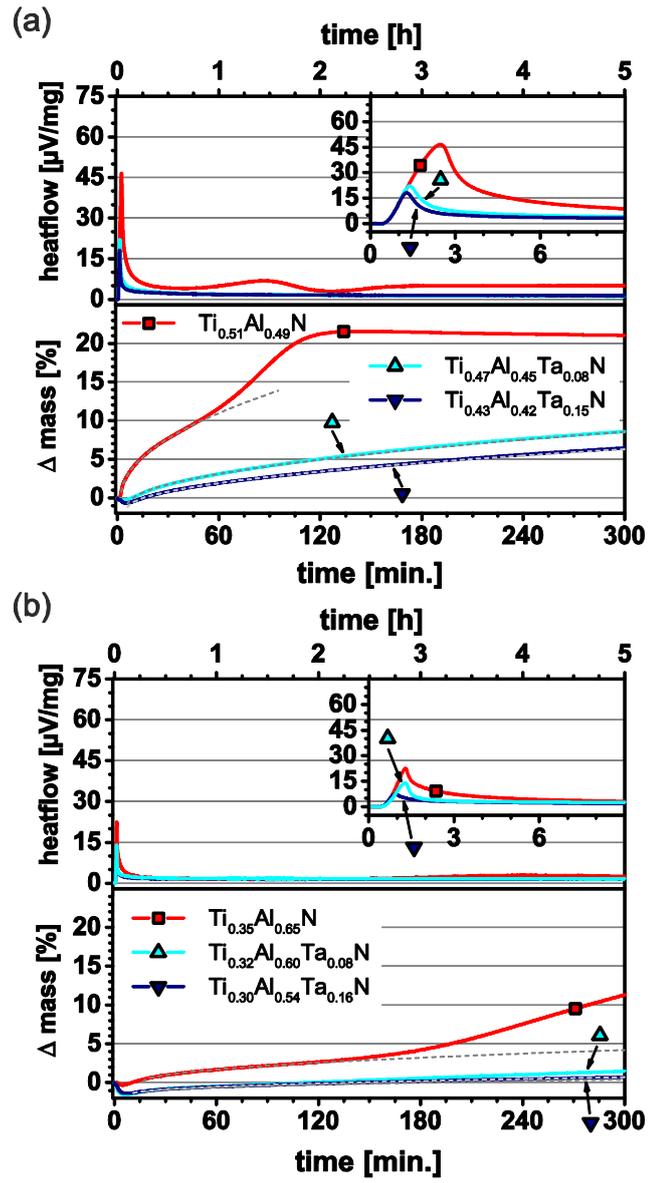

Fig. 5

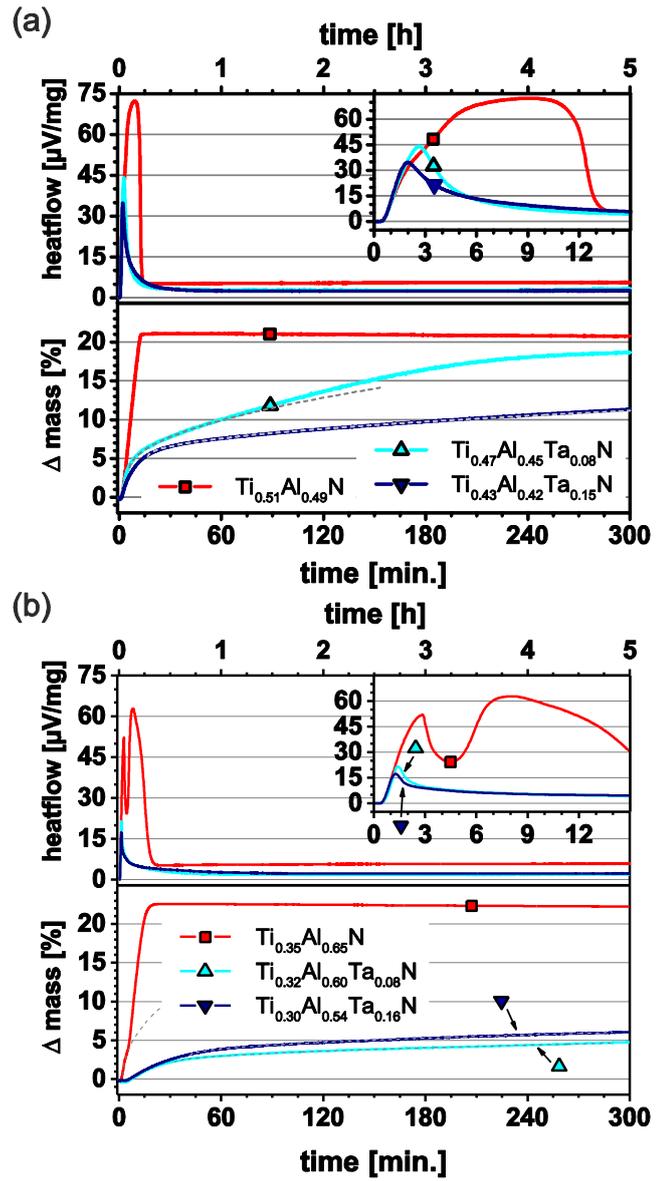

Fig. 6

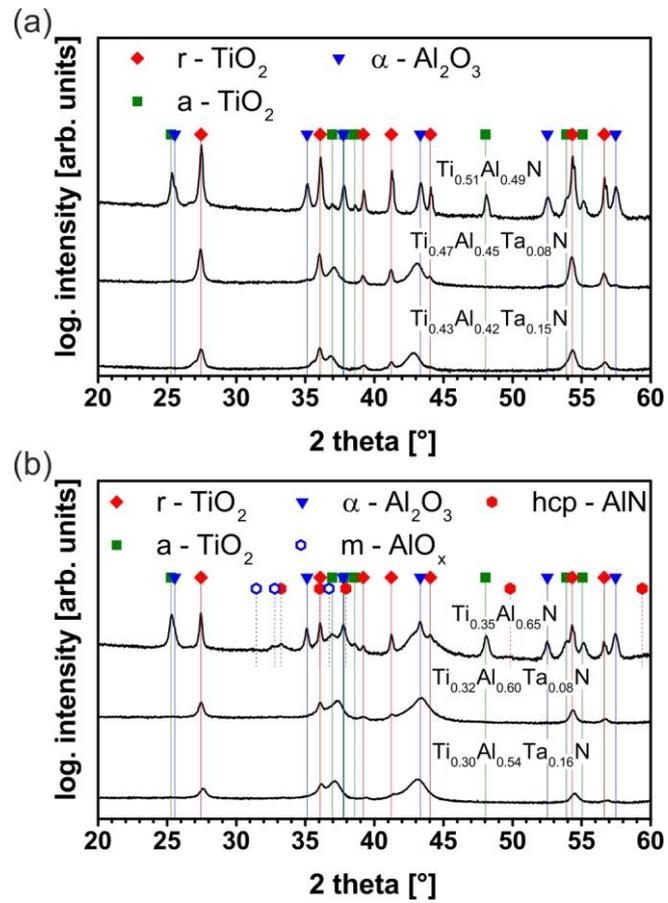

Fig. 7

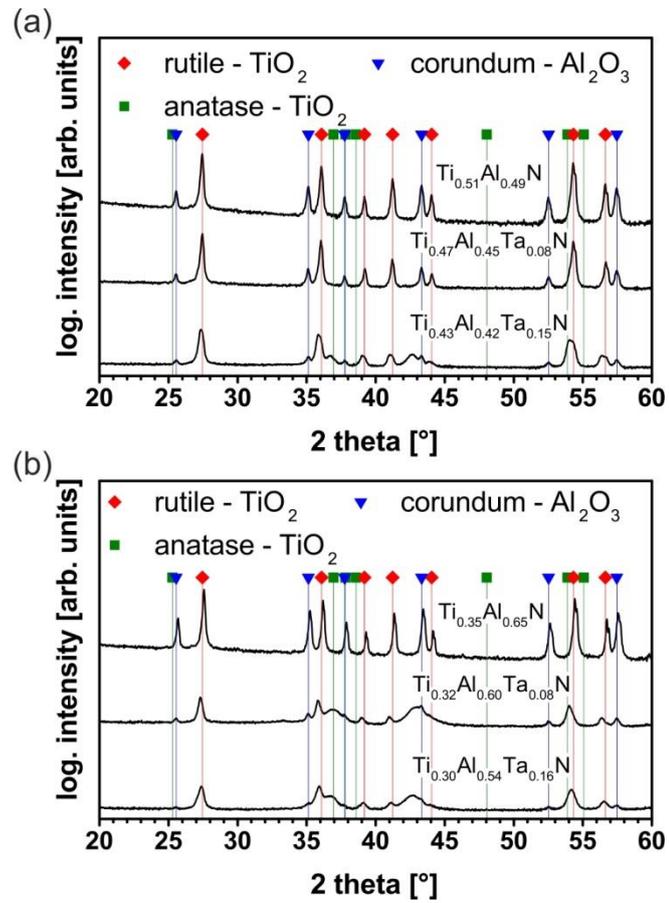

Fig. 8

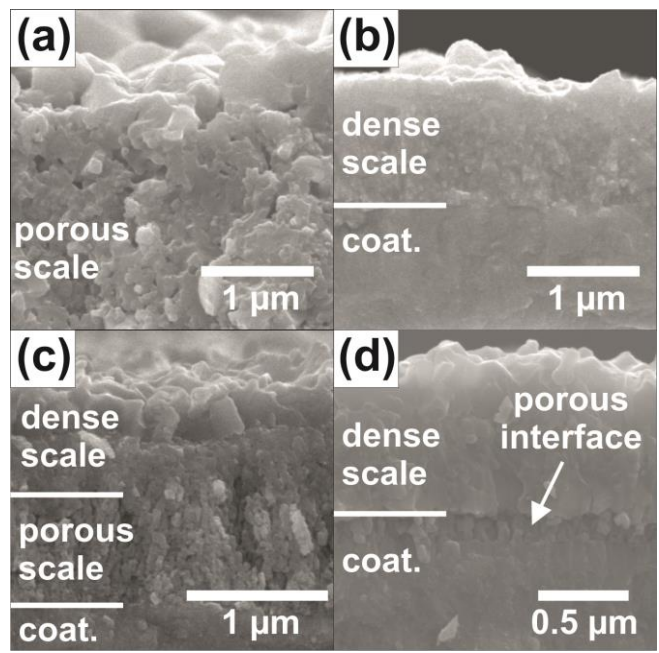

Fig. 9

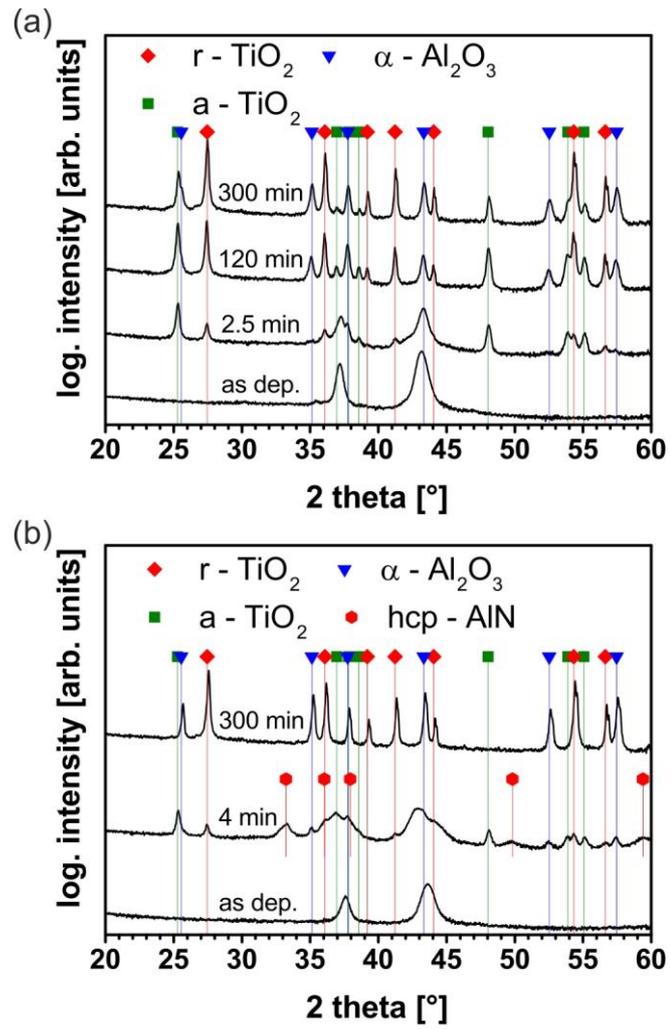

Fig. 10